\documentclass[12pt]{article}

\newcommand{\biblist}{\begin{list}{}
{\listparindent 0.0cm \leftmargin 0.50cm \itemindent -0.50 cm
\labelwidth 0 cm \labelsep 0.50 cm
\usecounter{list}}\clubpenalty4000\widowpenalty4000}
\newcommand{\ebiblist}{\end{list}}

\usepackage{bm}
\usepackage{natbib}
\usepackage{comment}
\usepackage{enumerate}  
\usepackage{hyperref}

\usepackage{latexsym}
\usepackage{amsmath}

\newtheorem{theorem}{Theorem}

\usepackage{color,graphicx,lscape,longtable}

\def\boxit#1{\vbox{\hrule\hbox{\vrule\kern6pt\vbox{\kern6pt#1\kern6pt}\kern6pt\vrule}\hrule}}

\newtheorem{lemma}{Lemma}[section]

\begin{document}

\title{Semiparametric response model with nonignorable nonresponse
}
\author{MASATOSHI UEHARA \\
Department of Statistics, Harvard University  \\ \href{a}{uehara\_m@g.harvard.edu}
\and JAE KWANG KIM \\
Department of Statistics, Iowa State University   \\
\href{a}{jkim@iastate.edu}
}

\baselineskip .3in
\maketitle 

\begin{abstract}
How to deal with nonignorable response is often a challenging problem encountered in statistical analysis with missing data. Parametric model assumption for the response mechanism is often made and there is no way to validate the model assumption with missing data. We consider a semiparametric response model that relaxes the parametric model assumption in the response mechanism. Two types of efficient estimators, profile maximum likelihood estimator and profile calibration estimator, are proposed and their asymptotic properties are investigated. Two extensive simulation studies are used to compare with some existing methods. We present an application of our method using Korean Labor and Income Panel Survey data.
\end{abstract}

\newpage

\section{Introduction}

Statistical analysis with missing data is an area under extensive research in recent years. To analyze partially missing data, the first step is to understand the response mechanism that  causes  missingness in the data. 
If the missingness for the study variable is conditionally independent of that variable, conditional on the other auxiliary variables, the response mechanism is called missing at random or ignorable, in the sense of \cite{rubin76}.  Otherwise, the response mechanism is called missing not at random or nonignorable.  It is more challenging to handle nonignorable nonresponse because  the assumed response model  cannot be
verified from  the observed study variables only and the response model may not be identifiable. Some identification assumptions are often made for the response model for the sake of making a valid inference from the incomplete data. Furthermore, the parametric model approach is known to be sensitive to the failure of the assumed parametric model \citep{kenward98}.  To obtain a robust result, it is desirable  to make the weakest possible model assumptions on the response mechanism. \cite{KimJaeKwang2013SMfH} contains a comprehensive review of the methods for parameter estimation under nonignorable nonresponse.

Instead of making parametric model assumptions for the response mechanism, we consider a semiparametric response model which allows more flexibility for the response mechanism. The semiparametric response model was first considered in Kim and Yu (2011), but the proposed method requires a validation sample for estimating model parameters. 
Shao and Wang (2016) also considered the same semiparametric model and proposed a parameter estimation method based on a calibration approach (Kott and Chang, 2008).  The proposed method of Shao and Wang (2016) is not necessarily efficient. We consider more efficient parameter estimation methods under the same semiparametric response model. 

We consider the following goals to meet in this paper. First of all, by taking the profile maximum likelihood estimation and profile calibration approach, we propose more efficient estimators under the semiparametric response model than the previous proposed estimator in \cite{shao16}. 
Next, we also propose a more efficient estimator for mean functional than the inverse probability weighted estimator used in \cite{shao16}. However, the proposed method can be computationally heavy as it involves another nonparametric estimator for the outcome model in addition to the nonparametric part in the response model. To solve this problem, we propose an approach using a parametric working model for the outcome model in the same spirit of the generalized estimating equation \citep{LiangKung-Yee1986LDAU}. Under such approach, when the working model is well-specified, the asymptotic variance form is not changed compared with an estimator when using a nonparametric model for the outcome model. Additionally, even when the working model is mis-specified, the estimators for the response probability and mean functional are still consistent. 

The paper is organized as follows. In Section 2, the basic setup and the model are introduced. In Section 3, the proposed method for estimating the parameters in the semiparametric response model is presented. In Section 4, estimation of mean functional using the proposed method is discussed. In Section 5, results from two extensive simulation studies are presented. In Section 6, we present an application of the proposed method  using Korean Labor and Income Panel Survey data. 
Some concluding remarks are made in Section 7. 

\section{Basic setup}

Let $(X,Y)$ be a vector of random variables from $F$, completely unspecified, and $Y$ is subject to missingness. We are interested in estimating $\mu=E(Y)$ from the $n$ independent  observations of  $(x_i, \delta_i, \delta_i y_i)$, $i=1, \cdots, n$, where $\delta_i$ is the response indicator function of $y_{i}$, that is, $\delta_i=1$ if $y_i$ is observed and $\delta_i=0$ otherwise.  If the response mechanism is ignorable in the sense that 
\begin{equation}
P ( \delta = 1 \mid x, y) = P( \delta =1 \mid x) ,
\label{1}
\end{equation}
then parameter estimation in the model $f(y \mid x)$ can be made without an explicit model assumption for the response mechanism. If assumption (\ref{1}) is believed to be unrealistic, then we often make a strong model assumption on $P( \delta =1 \mid x, y)$, such as 
\begin{equation}
P( \delta =1 \mid x, y) = \pi(x,y; \phi) 
\label{2}
\end{equation}
for some known function $\pi ( \cdot) \in (0,1]$ with unknown parameter $\phi$. For example, the logistic regression model 
\begin{equation}
\label{eq:para}
\pi(x,y; \phi) = \frac{ \exp ( \phi_0 + \phi_1 x + \phi_2 y) }{ 1+  \exp ( \phi_0 + \phi_1 x + \phi_2 y)  } 
\end{equation}
can be used in the response model (\ref{2}). Estimating the model parameter in (\ref{2}) is challenging because the model can be non-identifiable. A sufficient condition for model identification is to assume that $x=(x_1, x_2)$ and $x_2$ is conditionally independent of $\delta$ given $(x_1, y)$. Variable $x_2$ is called nonresponse instrumental variable \citep{wang14}.

Under the correct specification of the parametric response model and instrumental variable assumption, several methods are available for estimating the parameter $\phi$. There are mainly four approaches: (1) maximum likelihood estimation approach \citep{riddles16,morikawa17}, (2) calibration approach based on Generalized Method of Moments \citep{chang08,shao16,morikawa17_2}, (3) empirical likelihood approach \citep{qin02,TangNiansheng2014ELFE}, (4) pesudo likelihood approach \citep{tang03,zhao15}.  Note that (2) and (3) are essentially the same \citep{MorikawaKosuke2018Anot,QinJing2017BSOP}. Elaborating on the first and the second approaches, we propose estimation methods under the following semiparametric response model, which is a more general class than the parametric response model. The semiparametric response model is represented as 
\begin{align}
\pi(x,y)= \frac{\exp\{g(x)-\gamma y\}}{1+\exp \{ g(x)-\gamma y\}},
\label{eq:semi}
\end{align}
where $g( \cdot)$ is completely unspecified. This model was firstly introduced in \cite{Kim11b}. 

There are two things to note about this model. First, this is a more flexible model compared with commonly used parametric logistic models such as \eqref{eq:para}. Secondly, it is also viewed as a natural extension of the nonparametric response model in the ignorable case \citep{hirano03} by fixing $\gamma=0$. Although this semiparametric response model has the above good properties, it was required that  $\gamma$ is either known or estimable from a validation sample.  \cite{shao16} introduced an estimation method without using a validation sample by assuming the existence of $x_2$ in $x=(x_1, x_2)$, which is the nonresponse instrumental variable. Under this assumption, $g(x)$  becomes a function of $x_{1}$. We also assume this assumption for handling the identification problem of $\gamma$.

\section{Estimation under semiparametric response model}

Under the  semiparametric response model \eqref{eq:semi} and nonresponse instrumental variable assumption,  we propose new methods for estimating $g(x_{1})$ and $\gamma$. We denote the true $g(x_{1})$ as $g^{*}(x_{1})$ and the true $\gamma$ as $\gamma^{*}$ for clarity. We also write the evaluating function of $\gamma$ and $g$ at $\gamma^{*}$ and  $g^{*}$ as $\cdot |_{\gamma^{*}}$ and $\cdot |_{g^{*}}$. 

Under the semiparametric response  model in \eqref{eq:semi}, we can obtain
$$  E\bigg\{ \frac{ \delta}{ \pi (X_{1},Y) } -1 \mid X_{1}=x_1 \bigg \} \mid _{g^{*},\gamma^{*}} = 0 , $$
which leads to 
\begin{align*}
\exp\{ g_{\gamma}(x_1) \} = \frac{  E\big\{\delta \exp ( \gamma Y ) \mid  x_1\big\} }{  E\big \{1-\delta \mid x_1 \big\}}, \, \exp\{g^{*}(x_{1})\} =\exp\{ g_{\gamma}(x_1) \} \mid _{\gamma^{*}}.
\end{align*}
For fixed $\gamma$, we can use nonparametric methods to estimate $\exp\{ g_{\gamma}(x_1) \}$. More specifically, when the sample space is continuous, we can use a Kernel regression estimator
$$ \exp \{ \hat{g}_{\gamma} (x_1) \} = \frac{ \sum_{i=1}^n  \delta_i \exp ( \gamma y_i) K_h ( x_1 - x_{1i}) }{ \sum_{i=1}^n (1-\delta_i) K_h (x_1 - x_{1i} ) },$$ 
where $K_{h}$ is a kernel with bandwidth $h$. 

When the sample space is discrete, we can use 
\begin{align*}
    \exp \{ \hat{g}_{\gamma} (x_1) \} = \frac{ \sum_{i=1}^n  \delta_i \exp ( \gamma y_i) \mathrm{I}(x_1=x_{1i})}{ \sum_{i=1}^n (1-\delta_i) \mathrm{I}(x_1=x_{1i})},
\end{align*}
where $\mathrm{I}(\cdot)$ denotes an indicator function. For the convenience of notation, whether the sample space is continuous or not, we denote them as 
\begin{align*}
\exp \{ \hat{g}_{\gamma} (x_1) \} =  \frac{ \tilde{ E}\big \{ \delta \exp ( \gamma Y ) \mid x_1\big \}}{ \tilde{ E}\{1-\delta \mid x_1 \}}.
\end{align*}
Then, as discussed in \cite{shao16}, the profile response probability is obtained:
$$ {\pi}_p (x_{1i}, y_i;  \gamma ) = \frac{ \exp\{ \hat{g}_{\gamma}(x_{1i}) -\gamma y_i  \} }{ 1+ \exp\{ \hat{g}_{\gamma}(x_{1i})-\gamma  y_i  \} } .  $$

Using the above profile response probability, we take two approaches for the estimation of $\gamma$: (1) maximum likelihood estimation approach, (2) calibration approach. In both cases, the idea is to use $\pi_{p}(x)$ in the objective function, which is considered to be suitable when $g(x_{1})$ is known, in the same spirit of the profile likelihood approach in \cite{MurphyS.A.2000OPL}. Note that once $\gamma$ is estimated, we can use $\pi_{p}(x_{i},y_{i};\hat{\gamma})$ as the estimated profile response probability. Here, we first explain the maximum likelihood estimation approach and the calibration approach later. 

\subsection{Maximum likelihood estimation approach}

To compute the  maximum likelihood estimator in the missing data setting, we usually take two directions: (1) maximizing observed likelihood, (2) solving mean score equation. We start with the second direction. In appendix, we show that the estimator derived from the second perspective can also be interpreted from the first perspective. 

When $g(x_{1})$ is known, the observed score equation is
 \begin{align*}
        0 = \sum_{i=1}^{n}\bigg [ \delta_{i}\Big \{1-\pi(x_{1i},y_{i};g^{*},\gamma)\Big \}y_{i}-(1-\delta_{i}) E_{0}\big \{\pi(X,Y;g^{*},\gamma) Y \mid x_{i}\big \}\bigg ],
\end{align*}
where $ E_{0}[\cdot]$ denotes the expectation with respect to the distribution of $y$ given $x$ among $\delta=0$. 
By replacing $g(x_{1})$ with $\hat{g}_{\gamma}(x)$ and $ E_{0}[\pi y \mid x_{i}]$ with a nonparametric estimator, the estimator $\hat{\gamma}_{p-score}$ is defined as the solution to
\begin{align}
\label{eq:ms}
       0 = \sum_{i=1}^{n}\bigg [\delta_{i}\Big \{1-\pi_{p}(x_{1i},y_{i};\gamma)\Big \}y_{i}-(1-\delta_{i}) \frac{\tilde{ E}\big \{\delta\exp(\gamma Y)\pi_{p}(X_{1},Y;\gamma))Y \mid x_{i}\big \}}{\tilde{ E}\big \{\delta\exp(\gamma Y) \mid x_{i}\big \}}\bigg ]. 
\end{align}
This is based on the following relationship \citep{Kim11b}:
\begin{align*}
     E_{0}\big \{e(X,Y) \mid x\big \}=\frac{ E_{1}\big\{\exp(\gamma Y)e(X,Y) \mid x\big \}}{ E_{1}\big \{\exp(\gamma Y) \mid x\big \}},
\end{align*}
where $e(x,y)$ is any function of $(x,y)$. 

We now derive asymptotic results of the estimator obtained from \eqref{eq:ms}. By imposing technical conditions, we can ensure consistency by following Theorem 5.11 in \cite{BolthausenErwin2002LoPT}. For detail, see appendix. We obtain the following lemma as a first step. This lemma is also important in the next subsection for developing a new estimator from the calibration approach.

\begin{lemma}
\label{score_lemma}
We define the right hand side of \eqref{eq:ms} as $t(\mathbf{w})$. Under certain conditions, we have 
\begin{align*}
    \frac{1}{n}t(\mathbf{w})=\frac{1}{n}\sum_{i=1}^{n}\bigg \{\frac{\delta_{i}}{\pi_{p}(x_{1i},y_{i};\gamma)}-1\bigg \}\frac{ E\big \{\delta \exp(\gamma Y)\pi_{p}(X,Y;\gamma) \mid x\big \}}{ E\big \{\delta \exp(\gamma Y) \mid x\big \}}+\mathrm{o}_{p}(n^{-1/2}). 
\end{align*}
\end{lemma}
Based on this lemma, we obtain the following result. 
\begin{theorem}
\label{thm:score}
The asymptotic variance of $\hat{\gamma}_{p-score}$ is $n^{-1}A_{1}^{-1}B_{1}A_{1}^{-1}$ where
\begin{align*}
    A_{1} &=  E\left[O(X,Y) \pi(X,Y)(Y- E_{0}[Y \mid X_{1}])\big \{ E_{0}(\pi Y \mid X)- E_{0}(\pi Y \mid X_{1})\big \}\right ] \mid _{g^{*},\gamma^{*}}, \\
    B_{1} &=  E\left[O(X,Y) \big \{ E_{0}(\pi Y \mid X)- E_{0}(\pi Y \mid X_{1})\big \}^{2} \right] \mid _{g^{*},\gamma^{*}},\\
  \text{and} \\ 
    O(X,Y) &= \frac{1-\pi(X,Y)}{\pi(X,Y)}=\exp\big \{-g(X_{1})+\gamma Y\big \}.
\end{align*}
\end{theorem}

In practice, the estimator $\hat{\gamma}_{p-score}$ is unstable because of the double use of kernel estimators. Instead of using a kernel estimator for the calculation of the conditional expectation $ E_{0}(\pi y \mid x_{i})$, one can also use a model $f(y \mid x,\delta=1;\beta)$ for a density $y$ given $x$ and $\delta=1$ and apply fractional imputation \citep{Kim11}. We can replace the conditional expectation term in \eqref{eq:ms} with 
\begin{align*}
    \frac{\sum_{j=1}^{s}\exp(\gamma y_{ij})\pi_{p}(x_{1i},y_{ij};\gamma)y_{ij}}{\sum_{j=1}^{s}\exp(\gamma y_{ij})},
\end{align*}
where $\{y_{ij}\}_{j=1}^{s}$ is a $s$-size sample obtained from the conditional density $f(y \mid x,\delta=1;\hat{\beta})$ and $\hat{\beta}$ is the maximum likelihood estimator based on the observed samples. Throughout our work, we call the conditional distribution of $Y$ given $X$ and $\delta=1$ as an outcome distribution and $f(y \mid x,\delta=1;\beta)$ as a working parametric model. To avoid confusion, we write the estimator using this working parametric model as $\hat{\gamma}_{pw-score}$. 

\subsection{Calibration approach}

Here, we consider the calibration approach \citep{chang08}. Based on this approach, when $X_{2}$ is a discrete random variable taking values from $1$ to $l$,  \cite{shao16} introduced an estimator from the following moment conditions
\begin{equation}
\label{eq:cali1}
\sum_{i=1}^n \left\{ \frac{\delta_i}{  {\pi}_p (x_{1i}, y_i; \gamma )  } -1  \right\}v(x_{2i})= 0,
\end{equation}
where $v(x_{2})=\left(\mathrm{I}(x_{2}=1),\cdots,\mathrm{I}(x_{2}=l)\right)^{\top}$, using Generalized Method of Moments \citep{Hansen82} . 

Although this estimator does not require a validation sample for estimating $\gamma$, its performance is unstable because of the poor choice of  $v(x_{2i})$ in the the moment condition. We consider a broader class of estimators based on the following moment conditions:
\begin{equation}
\label{eq:cali_gene}
\sum_{i=1}^n \left\{ \frac{\delta_i}{  {\pi}_p (x_{1i}, y_i; \gamma) } -1  \right\} m(x_i;\gamma)=0,
\end{equation}
where $m(x;\gamma)$ is a function vector of $x$ including parameter $\gamma$. We call this estimator $\hat{\gamma}_{p-gmm}$.

Here, the problem is how to choose the control variable $m(x;\gamma)$ in \eqref{eq:cali_gene}. We can easily deduce that the efficiency will increase as we increase the dimension of $m(x;\gamma)$. However, for the current problem, the response probability $\pi_{p}$ includes a kernel estimator; thus, the calibration estimator \eqref{eq:cali_gene} from using a high-dimensional $m(x;\gamma)$ is computationally heavy. Thus, we suggest two one-dimensional moment conditions by deriving an asymptotic result of the estimator based on \eqref{eq:cali_gene}.

The asymptotic result of the calibration estimator from \eqref{eq:cali_gene} is given in the following theorem. 

\begin{theorem}
\label{thm:pm-gmm}
The asymptotic variance of the estimator $\hat{\gamma}_{p-gmm}$ is $n^{-1}A_{m}^{-1}B_{m}A_{m}^{-1}$:
\begin{align*}
    A_{m} &=  E\left(O(X,Y)\pi(X,Y) \big \{Y- E_{0}(Y \mid X_{1})\big\}\big [m(X;\gamma)- E_{0}\big \{m(X;\gamma) \mid X_{1}\big \}\big]\right) \mid _{g^{*},\gamma^{*}},\\
    B_{m} &=   E\left (O(X,Y)\left[m(X;\gamma)- E_{0}\big \{m(X;\gamma) \mid X_{1}\big \}\right]^{2}\right) \mid _{g^{*},\gamma^{*}}.
\end{align*}
\end{theorem}

There are two things to note. First, when $g(x_{1})$ is known, the term $ E_{0}[Y \mid X_{1}]$ and $ E_{0}[m(X) \mid X_{1}]$ will vanish. Second, a consistent estimator for the asymptotic variance, which can be used to construct a confidence interval, is derived as $\hat{A}_{m}^{-1}\hat{B}_{m}\hat{A}_{m}^{-1}$:
\begin{align*}
    \hat{A}_{m} &= \frac{1}{n}\sum_{i=1}^{n}\frac{1-\hat{\pi}_{i}}{\hat{\pi}_{i}}\delta_{i}\left \{y_{i}-\tilde{ E}_{0}(Y \mid x_{1i};\hat{\gamma})\right\}\left\{m(x_{i})-\tilde{ E}_{0}\big \{m(X) \mid x_{1i};\hat{\gamma}\big \}\right\},\\
    \hat{B}_{m} &= \frac{1}{n}\sum_{i=1}^{n}\frac{1-\hat{\pi}_{i}}{\hat{\pi}_{i}}\left [m(x_{i})-\tilde{ E}_{0}\{m(X) \mid x_{1i};\hat{\gamma}\}\right]^{2},
\end{align*}
where $\hat{\pi}_{i}=\pi_{p}(x_{i},y_{i};\hat{\gamma})$ denotes an estimated response probability. 

Based on the result of Theorem \ref{thm:pm-gmm}, two choices of one-dimensional $m(x;\phi)$ can be suggested. The first choice is the function appearing in Lemma \ref{score_lemma}. This function is derived by maximum likelihood estimation; thus, we can expect a good performance. Specifically, the estimator $\hat{\gamma}_{p-ca1}$ is defined as the solution to \eqref{eq:cali_gene} using
\begin{align*}
m(x;\gamma)=\frac{\mathrm{\tilde{E}}\big\{\delta \exp(\gamma Y)\pi_{p}(X_{1},Y;\gamma)Y \mid x\big \}}{\mathrm{\tilde{E}}\big \{\delta \exp(\gamma Y) |x\big \}},
\end{align*}
which is an approximation of the term in Lemma \ref{score_lemma}.

The second choice of $m(x;\gamma)$ is 
\begin{align*}
\frac{\mathrm{\tilde{E}}\big \{\delta \exp(\gamma Y)Y |x\big \}}{\mathrm{\tilde{E}}\big \{\delta  \exp(\gamma Y)/\pi_{p}(X_{1},Y;\gamma) |x\big \}}.
\end{align*}
We call the estimator from this $m(x;\gamma)$ as $\hat{\gamma}_{p-ca2}$. This is an approximation of $ E\big\{O(X,Y)\pi Y \mid X\big \}/ E\big \{O(X,Y) \mid X\big \}$. This choice is based on an optimal $m(x)$ if there are no $ E_{0}\big \{m(X) \mid X_{1}\big \}$ and $ E_{0}(Y \mid X_{1})$, that is, $g(x_{1})$ is known. This result is already known in other literature from different perspectives \citep{RotnitzkyAndrea1997AOSR, morikawa17_2}. We derive this result from a more direct approach. 

\begin{lemma}
\label{optimal}
When $g(x_{1})$ is known, the asymptotic variance is minimized when $m^{*}(X)= E\big \{O(X,Y)\pi Y \mid x\big \}/ E\big \{O(X,Y) \mid x\big \}$. 
\end{lemma}

Although $g(x_{1})$ is unknown in practice and the optimality does not hold in general, we can still consider this function as a good candidate of $m(x;\gamma)$. 

Finally, we derive asymptotic results of two estimators. 
\begin{theorem}
\label{thm:asympo_all}
The asymptotic variance of the estimator $\hat{\gamma}_{p-ca1}$ is the same as $n^{-1}A^{-1}_{1}B_{1}A^{-1}_{1}$, which is given in Theorem \ref{thm:score}, $\hat{\gamma}_{p-score}$. The asymptotic variance of $\hat{\gamma}_{p-ca2}$ is $n^{-1}A_{2}^{-1}B_{2}A_{2}^{-1}$.
\begin{align*}
    A_{2} &=  E\left\{O(X,Y)\pi \left\{Y- E_{0}(Y \mid X_{1})\right\}\left(\frac{ E\{O(X,Y)\pi Y \mid X\}}{ E\{O(X,Y) \mid X\}}- E_{0}\left[\frac{ E\{O(X,Y)\pi Y \mid X\}}{ E\{O(X,Y) \mid X\}} \mid X_{1}\right]\right)\right\} \mid _{g^{*},\gamma^{*}},\\
    B_{2} &=  E\left\{O(X,Y)\left(\frac{ E\{O(X,Y)\pi Y \mid X\}}{ E\{O(X,Y) \mid X\}}- E_{0}\left[\frac{ E\{O(X,Y)\pi Y \mid X\}}{ E\{O(X,Y) \mid X\}} \mid X_{1}\right]\right)^{2}\right\} \mid _{g^{*},\gamma^{*}}.
 \end{align*}
\end{theorem}
We can see that the asymptotic variance of $\hat{\gamma}_{p-score}$ and $\hat{\gamma}_{p-ca1}$ are the same.
This result is natural because $m(X)$ derived from the asymptotic analysis of $\hat{\gamma}_{p-score}$ is directly used for the construction of the estimator $\hat{\gamma}_{p-ca1}$. As for the comparison between $\hat{\gamma}_{p-ca1}$ and $\hat{\gamma}_{p-ca2}$, it is difficult to say which one is superior theoretically. In $\S$\,5, we experimentally confirm that the two estimators perform similarly. In addition, the asymptotic variances can be estimated like $\hat{\gamma}_{p-gmm}$. However, in the case of $\hat{\gamma}_{p-ca2}$, it might be difficult to estimate practically because of triply nested expectations.  

To avoid using kernels twice, we can use a parametric model for the density of $Y$ given $X$ and $\delta=1$ to calculate the conditional expectation. As in the previous section, write the estimators using $f(y \mid x,\delta=1;\hat{\beta})$ as $\hat{\gamma}_{pw-ca1}$ and $\hat{\gamma}_{pw-ca2}$.
For these estimators, the following asymptotic property holds. This result is similar to the property of  generalized estimating equation \citep{LiangKung-Yee1986LDAU}.

\begin{lemma}
\label{lemma:gamma}
When the working parametric model $f(y \mid x,\delta=1;\beta)$ is well-specified, the forms of the asymptotic variance of $\hat{\gamma}_{pw-ca1}$ and $\hat{\gamma}_{pw-ca2}$ are the same as in Theorem \ref{thm:asympo_all}. When the working parametric model $f(y \mid x,\delta=1;\beta)$ is mis-specified, $\hat{\gamma}_{pw-ca1}$ and $\hat{\gamma}_{pw-ca2}$ are $\sqrt{n}$-consistent. 
\end{lemma}

Note that $\hat{\gamma}_{pw-score}$ has a different asymptotic property compared with $\hat{\gamma}_{pw-ca1}$ and $\hat{\gamma}_{pw-ca2}$. When the parametric model is well-specified, the form of the asymptotic variance of $\hat{\gamma}_{p-score}$ will be changed from that of Theorem \ref{thm:score}. In addition, when the parametric model is mis-specified, $\hat{\gamma}_{p-score}$ is not consistent anymore. Therefore, $\hat{\gamma}_{pw-ca1}$ is considered to be more robust than $\hat{\gamma}_{pw-score}$ although asymptotic variances of $\hat{\gamma}_{p-ca1}$ and $\hat{\gamma}_{p-score}$ are the same. 

Between $\hat{\gamma}_{pw-ca1}$ and $\hat{\gamma}_{pw-ca2}$, it is difficult to say which one is superior theoretically in terms of statistical efficiency. We can state that $\hat{\gamma}_{pw-ca2}$ is computationally superior to $\hat{\gamma}_{pw-ca1}$ because if the parametric working model belongs to an exponential family, the function $m(x;\gamma)$ can be calculated analytically without relying on Monte Carlo integration \citep{morikawa17}. 

\section{Estimation of mean functional}

 We have discussed estimation of $\gamma$ so far. Here, we discuss estimation for the mean functional $\mu\equiv  E(Y)$. We can consider the following three estimators:
 \begin{align*}
      \hat{\mu}_{ipw} &= \frac{1}{n}\sum_{i=1}^{n}\frac{y_{i}}{\pi_{p}(x_{1i};\hat{\gamma})},\,
          \hat{\mu}_{mp} = \frac{1}{n}\sum_{i=1}^{n}\bigg [\delta_{i}y_{i}+(1-\delta_{i})\frac{\tilde{ E}\big \{\delta \exp(\hat{\gamma} Y) Y \mid x_{i}\big \}}{\tilde{ E}\big \{\delta \exp(\hat{\gamma} Y) \mid x_{i}\big \}}\bigg ],\\
       \text{end} \\
          \hat{\mu}_{db} &= \frac{1}{n}\sum_{i=1}^{n}\left[ \frac{\delta_{i}y_{i}}{\pi_{p}(x_{1i};\hat{\gamma})}+\left \{1-\frac{\delta_{i}}{\pi_{p}(x_{1i};\hat{\gamma})}\right\}\frac{\tilde{ E}\big \{\delta \exp(\hat{\gamma} Y) Y \mid x_{i}\big \}}{\tilde{ E}\big \{\delta \exp(\hat{\gamma} Y) \mid x_{i}\big \}}\right].
 \end{align*}

The estimator $\hat{\mu}_{ipw}$ was used in \cite{wang14}. However, if we consider the original motivation of the semiparametric response model first introduced in \cite{Kim11b}, it will be natural to use the estimator $\hat{\mu}_{mp}$. The estimator $\hat{\mu}_{db}$ is introduced using the analogy of doubly robust estimator in the ignorable case \citep{robins94}. Note that this estimator is different from other doubly robust form estimators in the nonignorable case \citep{miao16,morikawa17_2}. 

When $\hat{\gamma}$ is a $\sqrt{n}$-consistent estimator, the following asymptotic result can be established. 

\begin{theorem} 
\label{thm:mu}
Under certain regularity conditions, for estimators $\hat{\mu}_{ipw}$ , $\hat{\mu}_{mp}$ and $\hat{\mu}_{db}$, we have
\begin{align*}
    \hat{\mu}_{ipw}&=C_{1}+C_{2}+C_{3}+\mathrm{o}_{p}(n^{-1/2}),\\
    C_{1}&=\frac{1}{n}\sum_{i=1}^{n} E_{0}(Y \mid x_{1i}),\,C_{2}=\frac{1}{n}\sum_{i=1}^{n}\frac{\delta_{i}}{\pi_{i}}\left\{y_{i}- E_{0}(Y \mid x_{1i})\right \},\,
    C_{3}= H_{1}(\hat{\gamma}-\gamma^{*}),\\
    \hat{\mu}_{db}&=\hat{\mu}_{mp}+\mathrm{o}_{p}(n^{-1/2})=D_{1}+D_{2}+D_{3}+\mathrm{o}_{p}(n^{-1/2}), \\
   D_{1}&=\frac{1}{n}\sum_{i=1}^{n} E_{0}(Y \mid x_{i}),\,D_{2}=\frac{1}{n}\sum_{i=1}^{n}\frac{\delta_{i}}{\pi_{i}}\left\{y_{i}- E_{0}(Y \mid x_{i})\right \},\,D_{3}= H_{2}(\hat{\gamma}-\gamma^{*}),
\end{align*}
where 
\begin{align*}
H_{1}= E\left [(1-\pi)\left\{Y- E_{0}(Y \mid X_{1})\right\}^{2}\right] |_{g^{*},\gamma^{*}},\,
H_{2}= E\left[(1-\pi)\left\{Y- E_{0}(Y \mid X)\right\}^{2}\right ] \mid _{g^{*},\gamma^{*}}.
\end{align*}
\end{theorem}

There are three things to note.  First, we can see that the asymptotic variances of $\hat{\mu}_{mp}$ and $\hat{\mu}_{db}$ are the same. This result makes sense if you consider
the ignorable case, which has been well established in literature \citep{Rotnitzky14}. Second, the asymptotic variance of estimators $\hat{\mu}_{mp}$ and $\hat{\mu}_{db}$ are generally smaller than that of $\hat{\mu}_{ipw}$ because the estimators use more information in the conditional expectation of $\hat{\mu}_{mp}$ and $\hat{\mu}_{db}$. Third, the asymptotic variance can be estimated by taking the variance in the expression of Theorem \ref{thm:mu}.

We can also use the working parametric model for calculating the conditional expectation for  $ E_{0}(Y \mid X)$. 
We write the estimators as $\hat{\mu}_{w-mp}$ and $\hat{\mu}_{w-db}$. We have the following properties.

\begin{lemma}
\label{lemma:mu}
When the working parametric model $f(y \mid x,\delta=1; \beta)$ is well-specified, the forms of the asymptotic variance of $\hat{\mu}_{w-mp}$ and $\hat{\mu}_{w-db}$ are not changed from that of Theorem \ref{thm:mu}. Also, when the working parametric model is well-specified, $\hat{\mu}_{w-db}$ is consistent. 
\end{lemma}

This suggests that when the parametric outcome model is used, $\hat{\mu}_{w-db}$ is considered to be superior to $\hat{\mu}_{w-mp}$ because even if this model is mis-specified, $\hat{\mu}_{w-db}$ is consistent, while $\hat{\mu}_{w-mp}$ is not generally consistent. It is related to a doubly robust form estimators in nonignorable nonresponse cases \citep{MiaoWang2015IaDR,miao16} although our estimator $\hat{\mu}_{w-db}$ has a different form. They proposed doubly robust estimators in the sense that it is consistent even if the underlying baseline response model ($g(x_{1})$) or outcome model ($f(y \mid x,\delta=1)$) is mis-specified under the correct assumption of an odds ratio model ($\exp(\gamma Y)$). In our setup, the response model  $g(x_{1})$ is nonparamaetric, and therefore cannot be mis-specified. Our proposed estimator $\hat{\mu}_{w-db}$ is robust in the sense that the estimator is consistent under the correct assumption of the odds ratio model even if the outcome model is mis-specified. 

\section{Simulation Study}

We conducted two simulation studies to compare the performance of the estimators for $\gamma$ and $\mu$. For the estimators of $\gamma$, we compared $\hat{\gamma}_{p-gmm}$, $\hat{\gamma}_{p-score}$, $\hat{\gamma}_{p-ca1}$ and $\hat{\gamma}_{p-ca2}$ or $\hat{\gamma}_{p-gmm}$, $\hat{\gamma}_{pw-score}$, $\hat{\gamma}_{pw-ca1}$ and $\hat{\gamma}_{pw-ca2}$.
Note that the estimator $\hat{\gamma}_{p-gmm}$ with $m(x)=x$ corresponds to a baseline estimator proposed in \cite{wang14}. 
In the estimator $\hat{\gamma}_{p-gmm}$, we adopt $x_{1}$ as $m(X)$. For the estimation of mean functional $\mu$, we compared three estimators: $\hat{\mu}_{ipw}$, $\hat{\mu}_{mp}$ and $\hat{\mu}_{db}$ or $\hat{\mu}_{ipw}$, $\hat{\mu}_{w-mp}$ and $\hat{\mu}_{w-db}$.   
We consider simulations under three conditions: (1) $X$ and $Y$ are discrete, (2) $Y$ and $X_{2}$ are continuous and $X_{1}$ is discrete, and (3) $X$ and $Y$ are continuous. For the case of (3), see next section. In all cases, the parameter values under the missing data models were chosen so that the overall missing rate was about 30\%. 

\subsection{Case where \texorpdfstring{$X$}{a} and \texorpdfstring{$Y$}{a} are discrete} 

 Let $X_{1}$ be a categorical random variable taking values $\{0,1,2,3\}$ and Y be a binary variable taking values $\{0,1\}$ and Z be a binary variable taking value $\{0,1\}$. The random variable $X_{1}$ follows a multinomial distribution with probability $(0.25,0.25,0.25,0.25)$ and the random variable $X_{2}$, independent of $X_{1}$, follows a Bernoulli distribution with probability $0.5$. The random variable $Y$ follows a Bernoulli distribution with probability $1/\{1+\exp(1.3-(X_{1}-1.6)^{2}-1.5X_{2})\}$. The random variable $\delta$ follows a Bernoulli distribution as follows; M1: $\pi(X_{1},Y)=1/\{1+\exp(-\phi_{0}-\phi_{1}X_{1}+\gamma Y)\}$, where $(\phi_{0},\phi_{1},\gamma)=(0.2,0.8,0.6)$, M2: $\pi(X_{1},Y)=1/\{1+\exp(\phi_{0}+\phi_{1}X_{1}+\phi_{2}X_{1}^{2}+\gamma Y)\}$, where $(\phi_{0},\phi_{1},\phi_{2},\gamma)=(0.2,-0.4,0.7,0.6)$, and M3: $\pi(X_{1},Y)=1/\{1+\exp(\phi_{0}+\phi_{1}\sin(X_{1})+\gamma Y)\}$, where $(\phi_{0},\phi_{1},\gamma)=(-1.6,-0.8,0.6)$. 
The simulation is replicated $500$ times with two sample sizes; $1000$ and $4000$. Here, as all the variables are discrete, we use a nonparametric model for $y$ given $x$ and $\delta=1$. 

The result for the estimation of $\gamma$ and $\mu$ are presented in Table \ref{tab:case-b} and Table \ref{tab:case-b-mu} respectively. First, we can see that our proposed estimators perform better than the baseline estimator $\hat{\gamma}_{p-gmm}$ in terms of mean square errors. Next, it is seen that $\hat{\mu}_{mp}$ and $\hat{\mu}_{db}$ are superior to $\hat{\mu}_{ipw}$ in terms of efficiency, which is consistent with our theory in Theorem \ref{thm:mu}. Finally, it is confirmed that when the sample size is large ($n=4000$), mean square errors of $\hat{\gamma}_{p-score}$ and $\hat{\gamma}_{p-ca1}$ are almost the same. This matches to theoretical results in $\S$\, 3. It is also seen that mean square errors of $\hat{\gamma}_{p-ca1}$ and $\hat{\gamma}_{p-ca2}$ are almost the same.

\begin{table}[!]
    \centering
    \caption{Monte Carlo mean square error (MSE) and bias of $\gamma$ when $X$ and $Y$ are discrete}
    \begin{tabular}{c c c c c c c}
     Model & n &  & $\hat{\gamma}_{p-gmm}$&  $\hat{\gamma}_{p-score}$ & $\hat{\gamma}_{p-ca1}$ & $\hat{\gamma}_{p-ca2}$  \\
     $M_{1}$ &1000 & Bias& 0.09 & 0.01 & 0.04 & 0.03  \\
        & & MSE &  0.94 & 0.61 & 0.44 & 0.46  \\
    & 4000 & Bias&  0.00 & 0.01 & 0.01 & 0.01  \\
    & & MSE &0.16 & 0.10 & 0.10 & 0.10  \vspace{1mm}  \\
     $M_{2}$ &1000 & Bias& 0.08 & 0.02 & 0.01 & 0.01  \\
        & & MSE &  0.52 & 0.39 & 0.34 & 0.35  \\
    & 4000 & Bias&  0.00 & 0.00 & 0.00 & 0.00  \\
    & & MSE &0.11 & 0.086 & 0.088 & 0.088       \vspace{1mm}\\
     $M_{3}$ & 1000 & Bias & 0.07 & 0.02 & 0.01 & 0.01  \\
        & & MSE &  0.43 & 0.35 & 0.32 & 0.33  \\
    & 4000 & Bias &  0.02 & 0.00 & 0.00 & 0.00  \\
    &     & MSE &  0.090 & 0.072 & 0.072 & 0.072  
    \end{tabular}
    \label{tab:case-b}
\end{table}

\begin{table}[!]
    \centering
    \caption{Monte Carlo mean square error (MSE) for $\mu$ when $X$ and $Y$ are discrete. Multiplied by 1000.}
    \begin{tabular}{c c c c c c c}
     Model & n &  &$\hat{\gamma}_{p-gmm}$ & $\hat{\gamma}_{p-score}$  & $\hat{\gamma}_{p-ce1}$ & $\hat{\gamma}_{p-ce2}$  \\
    $M_{1}$ & $1000$ & $\hat{\mu}_{ipw}$ & 0.92 & 0.73 & 0.73 & 0.73  \\
    &  & $\hat{\mu}_{mp}$ & 0.89 & 0.73  & 0.73 & 0.73 \\
    &  & $\hat{\mu}_{db}$ & 0.89 & 0.73  & 0.73 & 0.73 \\
    & $4000$ & $\hat{\mu}_{ipw}$ & 0.32 & 0.24 & 0.24 & 0.24   \\
    &  & $\hat{\mu}_{mp}$  & 0.31 & 0.24  & 0.24 & 0.24 \\
    &  & $\hat{\mu}_{db}$ & 0.31  & 0.24  & 0.24 & 0.24       \vspace{1mm} \\
  $M_{2}$ & $1000$ & $\hat{\mu}_{ipw}$  & 1.33 & 1.04 & 1.01 & 1.01   \\
    &  & $\hat{\mu}_{mp}$ & 1.32 & 1.01  & 0.99 & 1.00 \\
    &  & $\hat{\mu}_{db}$ & 1.32 & 1.01  & 0.99 & 1.00 \\
    & $4000$ & $\hat{\mu}_{ipw}$  & 0.35 & 0.28 & 0.28 & 0.27   \\
    &  & $\hat{\mu}_{mp}$  &  0.34 & 0.27  & 0.27 & 0.27 \\
    &  & $\hat{\mu}_{db}$  &  0.34 & 0.27  & 0.37 & 0.27  \vspace{1mm} \\
$M_{3}$ & $1000$ & $\hat{\mu}_{ipw}$  & 129 & 128  & 128 & 129   \\
    &  & $\hat{\mu}_{mp}$  & 1.32 & 1.17  & 1.17 & 1.17 \\
    &  & $\hat{\mu}_{db}$   & 1.32 & 1.17  & 1.17 & 117 \\
    & $4000$ & $\hat{\mu}_{ipw}$  & 122 & 123 & 122 & 122   \\
    &  & $\hat{\mu}_{mp}$ & 0.27 & 0.24  & 0.24 & 0.24 \\
        &  & $\hat{\mu}_{db}$ & 0.27 & 0.24  & 0.24 & 0.24   \\
    \end{tabular}
    \label{tab:case-b-mu}
\end{table}

\subsection{Cases where \texorpdfstring{$X_{1}$}{a} is discrete and \texorpdfstring{$Y,X_{2}$}{a} are continuous}

Let $X_{1}$ be a binary random variable, whose distribution is described above, and $X_{2}$ be an uniform distribution $U[-1,1]$. As for the random variable $Y$, we make an assumption for the conditional distribution of $Y$ given $X$ and $\delta=1$, and also $\pi(X,Y)=P(\delta=1|X,Y)$. Note that the distribution of $Y$ is uniquely determined by these two distributions. For the generation of samples in this setting, see \cite{morikawa17}. 
First, let the distribution of $Y$ given $X$ and $\delta=1$ be a Gaussian distribution $\mathrm{N}(-1.0-0.4X_{1}+0.5X_{2}^{2},1.0)$
. Second, let the response mechanism be the following logistic models; M1: $\pi(X_{1},Y)=1/\{1+\exp(-\phi_{0}-\phi_{1}X_{1}+\gamma Y)\}$, where $(\phi_{0},\phi_{1},\gamma)=(0.3,0.4,0.5)$, M2: $\pi(X_{1},Y)=1/\{1+\exp(-\phi_{0}-\phi_{1}X_{1}-\phi_{2}X_{1}^{2}+\gamma Y)\}$, where $(\phi_{0},\phi_{1},\phi_{2},\gamma)=(0.3,0.3,0.2,0.5)$, and M3: $\pi(X_{1},Y)=1/\{1+\exp(-\phi_{0}-\phi_{1}\sin(X_{1})+\gamma Y)\}$, where $(\phi_{0},\phi_{1},\gamma)=(0.3,0.3,0.5)$. 

We used a well-specified parametric outcome model to assist with the calculation of the conditional expectation and compared $\hat{\gamma}_{p-gmm}$, $\hat{\gamma}_{pw-score},\hat{\gamma}_{pw-ca1}$ and $\hat{\gamma}_{pw-ca2}$.  Note that the estimator for $\exp\big \{-g(x_{1})\big \}$ is still constructed from an empirical distribution without using a kernel estimator. For the calculation of the condition expectation term based on Monte Carlo integration, which appears in the objective function of $\hat{\gamma}_{pw-score}$, $\hat{\gamma}_{pw-ca1}$ and $\hat{\gamma}_{pw-ca2}$, we obtain $500$ samples from the auxiliary distribution $f(y \mid x,\delta=1;\hat{\beta})$. However, in the case of $\hat{\gamma}_{pw-ca2}$, actually, we can perform calculation analytically because $f(y \mid x,\delta=1;\beta)$ is a normal distribution \citep{morikawa17_2}. We write the estimator relying on Monte Carlo integration as $\hat{\gamma}_{pw-ca2-s}$ and the other that does not  rely on Monte Carlo integration as $\hat{\gamma}_{pw-ca2-a}$. 

The simulation is replicated $500$ times with sample size $2000$ and $4000$. The result is reported in Table \ref{tab:case-c} and \ref{tab:case-c-mu}. There are five things to note. First, it is seen that all of the proposed estimators are superior to the baseline estimator $\hat{\gamma}_{p-gmm}$.  Secondly, whether $\hat{\gamma}_{p-ca1}$  or $\hat{\gamma}_{p-ca2}$ is better depends on the true distribution.  Thirdly,  the mean square error of $\hat{\gamma}_{p-score}$  is smaller than that of $\hat{\gamma}_{p-ca1}$   and $\hat{\gamma}_{p-ca2}$.  However, as noted earlier,  $\hat{\gamma}_{p-score}$ is not a consistent estimator when the working model is mis-specified. 
Fourth, by comparing two estimator $\hat{\gamma}_{p-ca2-s}$ and $\hat{\gamma}_{p-ca2-a}$, we can see the variance increases negligibly due to Monte Carlo integration. Finally, it is seen that $\hat{\mu}_{mp}$ and $\hat{\mu}_{db}$ are superior to $\hat{\mu}_{ipw}$ in terms of mean square errors. 

\begin{table}[!]
    \centering
    \caption{Monte Carlo mean square error (MSE) and bias of $\gamma$ when $X_{1}$ is discrete, and $X_{2},Y$ are continuous}
    \begin{tabular}{c c c c c c c c}
  Model & n & & $\hat{\gamma}_{p-gmm}$ & $\hat{\gamma}_{pw-score}$ & $\hat{\gamma}_{pw-ca1}$ & $\hat{\gamma}_{pw-ca2-a}$ & $\hat{\gamma}_{pw-ca2-s}$ \\
 $M_{1}$ & $2000$ & Bias & 0.54 & 0.05  & 0.03  & 0.01  & 0.01     \\
    	& 			& MSE   & 1.30     &  0.27  & 0.41  & 0.36   & 0.36  \\
    & $4000$ & Bias &  0.53  & 0.02  & 0.01  & 0.02 & 0.01 \\
    & 			& MSE   & 1.27    &  0.15  & 0.19  & 0.18   & 0.21  \vspace{1mm}  \\
$M_{2}$ & $2000$ & Bias & 0.61 & 0.05   & 0.07  & 0.08  & 0.09     \\
    & 			& MSE&  1.25 & 0.27 & 0.45 & 0.39 & 0.40   \\
    & $4000$ & Bias &  0.6 & 0.03 & 0.01 & 0.01 & 0.01  \\
    & 			& MSE  & 1.3 & 0.16 & 0.22 & 0.16 & 0.20 \vspace{1mm}  \\
 $M_{3}$ & $2000$ & Bias & 0.67  & 0.02  & 0.03  & 0.01 & 0.01   \\
    & 				& MSE & 1.31  & 0.24  & 0.43  & 0.38  & 0.38  \\
    & $4000$ & Bias &  0.62 & 0.03 & 0.02 & 0.01 & 0.01  \\
    & 				& MSE &  1.32 & 0.13 & 0.18 & 0.14 & 0.16 
    \end{tabular}
    \label{tab:case-c}
\end{table}

\begin{table}[!]
    \centering
    \caption{Monte Carlo mean square error (MSE) for $\mu$ when $X_{1}$ is discrete, and $X_{2},Y$ are continuous. Multiplied by 100. }
    \begin{tabular}{c c c c c c c}
     Model & n &  &$\hat{\gamma}_{p-gmm}$   & $\hat{\gamma}_{pw-score}$ & $\hat{\gamma}_{pw-ca1}$ & $\hat{\gamma}_{pw-ca2-a}$  \\
    $M_{1}$ & 2000 & $\hat{\mu}_{ipw}$ & 29 & 9.3 & 11.4 & 9.9   \\
    &  & $\hat{\mu}_{w-mp}$ & 11 & 2.4  & 3.8 & 3.5 \\
    &  & $\hat{\mu}_{w-db}$ & 10 & 2.2 & 3.7 & 3.4 \\
    & 4000 & $\hat{\mu}_{ipw}$ &  29 & 5.4 & 5.5 &  5.4  \\
    &  & $\hat{\mu}_{w-mp}$  & 11 & 1.4 &  1.6 & 1.7  \\
    &  & $\hat{\mu}_{w-db}$  & 11 & 1.3 & 1.6 & 1.7 \vspace{1mm}  \\
  $M_{2}$ & 2000 & $\hat{\mu}_{ipw}$  & 27 & 9.9 & 10 & 10   \\
    &  & $\hat{\mu}_{w-mp}$ &   13  & 2.2 & 3.7 & 3.4 \\
    & & $\hat{\mu}_{w-db}$ &   12  & 2.2 & 3.7 & 3.3 \\
    & 4000 & $\hat{\mu}_{ipw}$  & 26 &  3.0 & 6.0 &  4.8   \\
    &  & $\hat{\mu}_{w-mp}$  &  10 & 1.0 &  1.9 & 1.3 \\
    &  & $\hat{\mu}_{w-db}$   & 10 & 0.9 & 1.9 & 1.3 \vspace{1mm}  \\
    $M_{3}$ & 2000 & $\hat{\mu}_{ipw}$  &  35 & 8.5 & 12 & 12  \\
    &  & $\hat{\mu}_{w-mp}$  & 12 &  2.3  & 4.2 & 3.6 \\
    &  & $\hat{\mu}_{w-db}$  & 12 & 2.1  & 4.1 & 3.6 \\
    & 4000 & $\hat{\mu}_{ipw}$  & 20 & 6.1 & 5.9 & 5.8   \\
    &  & $\hat{\mu}_{w-mp}$   & 11 &  1.4 &  1.8 &  1.5\\
    &  & $\hat{\mu}_{w-db}$   & 11 & 1.3  & 1.7 & 1.5 \\
    \end{tabular}
    \label{tab:case-c-mu}
\end{table}

\subsection{Coverage probability}

We further examined the confidence intervals of $\gamma$ and $\mu$ under the setting $\S$\,5.1 with $n=4000$. As an estimator for $\mu$, we adopted $\hat{\mu}_{mp}$.   
We nonparametrically estimated the asymptotic variances based on the forms in the $\S$\,3 and $\S$\,4.
Note that our method for constructing confidence intervals does not rely on bootstrapping unlike \cite{shao16} because there is no guarantee that bootstrapping would work. The result is reported in Table \ref{tab:coverage} with the appropriate coverage rate. 

\begin{table}[!]
    \centering
    \caption{Coverage probability (CI) for $\gamma$ and $\mu$.}
    \begin{tabular}{c c c c c}
     Model &   & $\hat{\gamma}_{p-gmm}$ & $\hat{\gamma}_{p-ca1}$ & $\hat{\gamma}_{p-ca2}$  \\
     $M_{1}$ & $\gamma$  & 0.94 & 0.95 & 0.95 \\
    & $\mu$  & 0.94  & 0.95  & 0.94     \vspace{1mm}  \\
    $M_{2}$ & $\gamma$  & 0.93 & 0.94 & 0.95    \\
    & $\mu$  & 0.94  & 0.95  & 0.94 \vspace{1mm} \\
    $M_{3}$ & $\gamma$  & 0.95 & 0.95 & 0.96 \\
    & $\mu$  & 0.94 & 0.95 &  0.95  \\
    \end{tabular}
    \label{tab:coverage}
\end{table}

\section{Numerical illustration} 

We apply our method to the Korean Labor and Income Panel Survey data, which was used in several papers \citep{Kim11,shao16,morikawa17}.
These data include $n=2506$ Korean residents and four variables: the response variable $y$ is income ($10^{6}$ Korean Won) in the year, $x_{1}$: income in the previous year, $x_{2}$: gender, $x_{3}$: age, $x_{4}$: education level, where $x_{1}$ and $y$ are continuous variables, and $x_{2}$ has two categories, $x_{3}$ has three categories, and $x_{4}$ has two categories.  

We made an artificial incomplete dataset by assuming the two response models; M1: $\pi=1/\{1+\exp(- 1.3 -0.3\sqrt{x_{1}}-0.2x_{1}+0.6y)\}$ and M2: $\pi= 1/\{1+\exp(-1.2-0.5x_{1}+0.6y)\}$. Thus, under the assumption of $\pi=1/\{1+\exp(-g(x_{1})+\gamma y)\}$, we estimate $\gamma$ and $\mu$ and their confidence interval according to $\hat{\gamma}_{p-gmm}$, $\hat{\gamma}_{pw-ca1}$ and $\hat{\gamma}_{pw-ca2}$. As for $\hat{\gamma}_{p-gmm}$, we used $x_{2}$, $x_{3}$, $x_{4}$ as nonresponse instrumental variables.  For the estimation of $\mu$, we used $\hat{\mu}_{w-db}$. We used the bandwidth when regressing  $\delta$ on $X_{1}$ based on cross validation using the np R package \citep{npnp}.  

Note that the true mean income was $1.85$, which is calculated using the complete data. We reported an estimated value and confidence interval of $\mu$ for each case in Table \ref{tab:coverage2}. We can see that  $\hat{\gamma}_{p-ca1}$, $\hat{\gamma}_{p-ca2}$ are superior to $\hat{\gamma}_{p-gmm}$. 

\begin{table}[!]
    \centering
    \caption{Estimated value and confidence interval (CI) of $\hat{\mu}_{w-db}$ with significant level $\alpha=0.05$. The true mean is $1.85$.} 
    \begin{tabular}{c c c c c}
   Model & & $\hat{\gamma}_{p-gmm}$ &$\hat{\gamma}_{pw-ca1}$ &  $\hat{\gamma}_{pw-ca2}$ \\
  $M_{1}$ & Value & 1.80 & 1.86 & 1.86 \\
    & CI(0.05) & $\pm 0.12$ & $\pm 0.06$ &$\pm 0.07$ \vspace{1mm}  \\
 $M_{2}$ & Value &  1.77 & 1.84  &  1.85 \\
    & CI(0.05)  & $\pm 0.19$ & $\pm 0.08$  &$\pm 0.09$   \\
    \end{tabular}
    \label{tab:coverage2}
\end{table}

\section{Discussion}

We have developed several methods doing parameter estimation for the semiparametric response model. To summarize, we have the following practical suggestions. When the instrumental variables are discrete, we recommend using  calibration estimators $\hat{\gamma}_{p-ca1}$ or $\hat{\gamma}_{p-ca2}$ for the estimation of $\gamma$ by plugging in a nonparametric estimator in the conditional expectation term and then, estimating $\mu$ using the doubly robust form estimator $\hat{\mu}_{db}$. 
When the instrumental variables are continuous, we recommend using $\hat{\gamma}_{pw-ca2}$ for the estimation of $\gamma$ and using $\hat{\mu}_{w-db}$ for the estimation of $\mu$ with the aid of parametric working outcome model.

For future research, the following directions will be considered. The first is how to choose instrumental variables practically from available covariates. One can consider the following three step procedure: (a) making several models as in our work assuming each covariate is an instrumental variable, (b) performing estimation and calculating estimated likelihood, (c) select the model which maximizes the estimated likelihood.  The second direction is exploring more efficient estimators under the semiparametric response model or under the semiparametric response model and parametric outcome model. 
    
\newpage

\appendix

\section{Proof of results}

To derive asymptotic results, we assume the following conditions to use Lemma 8.11 in \cite{newey94} and Theorem 6.18. in \cite{BolthausenErwin2002LoPT}. Essentially, it requires three types of conditions: (1) stochastic equicontinuity, (2) the objective function can be linearized with respect to  nonparametric components (this also includes a convergence rate condition of nonparametric estimators), and (3) uniform convergence of the differentiation of estimating equation with respect to a parameter of interests. For simplicity, we write down required conditions for Theorem 1 when the sample space of $X$ is continuous. For other theorems and results where X is discrete, we can add conditions similarly. 

We define $\left\{\frac{\delta}{\pi(x,y;g,\gamma)}-1\right\}h(x;\gamma)$ as $\phi(w;g,\gamma)$. We denote the true $g(x_{1})$ as $g^{*}(x_{1})$ and the true $\gamma$ as $\gamma^{*}$. For the norm of functional space including $g(x_{1})$, we introduce a norm $\|\cdot \|$ in a H\"{o}lder space. 

(C1): The response probability is uniformly bounded below from zero. 

(C2): $W_{i}=(X_{i},Y_{i},R_{i})$ are independent and identically distributed.

(C3): The parameter space $\Gamma$ for $\gamma$ is compact.

(C4): The kernel $K(x)$ has bounded derivatives of order $k$, satisfies $\int K(x)\mathrm{d}x=1$, and has zero moments of order up to $m-1$ and nonzero $k$-th order moment. 

(C5): For all $y$, $\pi(\cdot,y)$ and $\exp\{g_{\gamma}(\cdot)\}$ are differentiable to order $k$ and are bounded on an open set containing support of $x$.

(C6): There exists $v\geq 4$ such that $E(|\delta \exp(\gamma Y)\mid ^{v})$ and $E(|1-\delta |^{v})$, and $E(|\delta \exp(\gamma Y)\mid ^{v}|x)f(x)$ and $E(|1-\delta |^{v}|x)f(x)$  are bounded for all supports of $x$. 

(C7): As $h\to 0$, $n^{1-2/v}h/\log n\to \infty$, $n^{1/2}h^{1+2k}/\log n\to \infty$ and $n^{1/2}h^{2k}\to 0$. 

(C8): $0=E\{\phi(w;g_{\gamma},\gamma)\}$ has a unique solution with respect to $\gamma$ in the interior of $\Gamma $. (This condition is needed for consistency. In that sense, it overlaps the condition below. )

(C9): Consistency:$(\hat{\gamma}_{p-gmm},\hat{g}_{\hat{\gamma}})\stackrel{p}{\rightarrow}(\gamma^{*},g^{*})$. This is discussed in Section C. 

(C10): The class of functions $\{\phi(w;g,\gamma):|\gamma-\gamma^{*}|<\delta, \|g-g^{*}\|<\delta \}$ is a Donsker class for some $\delta>0$. 

(C11): The map $\gamma \to E\{\phi(w;\gamma,g_{\gamma})\}$ is differentiable at $\gamma^{*}$, uniformly in $\gamma$ in a neighborhood of $\gamma^{*}$ with nonsigular derivative matrices. 

(C12): The map $(\gamma,g)\to\phi(w;\gamma,g)$ is continuous in $L^{2}$ space induced from a true distribution. 

\subsection{Proof of Theorem 2}

We have:
\begin{align}
0 &=  \frac{1}{n}\sum_{i=1}^{n}\left\{\frac{\delta_{i}}{\pi(x_{1i},y_{i};\hat{g}_{\gamma},\gamma)}-1\right\}m(x_{i})\nonumber \\
&= \frac{1}{n}\sum_{i=1}^{n}\left[\delta_{i}\{1+\exp(-\hat{g}_{\gamma}+\gamma y_{i})\}-1\right]m(x_{i}) \nonumber \\
&= \frac{1}{n}\sum_{i=1}^{n}\left[\delta_{i}\{1+\exp(-g_{\gamma}+\gamma y_{i})\}-1\right]m(x_{i})+\mathrm{Rm}(\mathbf{w}) \nonumber \\
&= \frac{1}{n}\sum_{i=1}^{n}\left\{\frac{\delta_{i}}{\pi(x_{1i},y_{i};g_{\gamma},\gamma)}-1\right\}m(x_{i})+\mathrm{Rm}(\mathbf{w})\label{eq:residual},
\end{align}
where the residual term $\mathrm{Rm}(\mathbf{w})$ is equal to
\begin{align*}
\frac{1}{n}\sum_{i=1}^{n}\delta_{i}\left[\exp\{-\hat{g}_{\gamma}(x_{1i})\}-\exp\{-g_{\gamma}(x_{1i})\}\right] m(x_{i}). 
\end{align*}
From the assumptions (C1)-(C8), we have 
\begin{align*}
\|E\{\delta \exp(\gamma y)\mid x\} f_{1}(x) -\mathrm{\tilde{E}}\{\delta \exp(\gamma y)\mid x\} f_{1}(x)\|&=\mathrm{o}_{p}(n^{-1/4}), \\
\|E(1-\delta|x)f_{1}(x)-\mathrm{\tilde{E}}(1-\delta|x)f_{1}(x)\| &=\mathrm{o}_{p}(n^{-1/4}),\\
\left \|\frac{\mathrm{\tilde{E}}\{\delta \exp(\gamma y)\mid x\}}{\mathrm{\tilde{E}}\{1-\delta|x\}}-\exp\{-g_{\gamma}(x)\}-\mathrm{Rm}_{2}(\mathbf{w})\right \|&=\mathrm{o}_{p}(n^{-1/2}),\\
\end{align*}
where
\begin{align*}
\mathrm{Rm}_{2}(\mathbf{w})=\frac{1}{n^{2}}\sum_{i=1}^{n}\sum_{j=1}^{n}\delta_{i}m(x_{i})\exp(\gamma y_{i})\mathrm{K}_{h}(x_{1i},x_{1j})\frac{1-\delta_{j}-\exp\{-g_{\gamma}(x_{i})\}\delta_{j} \exp(\gamma y_{j})}{E\{\delta \exp(\gamma Y)\mid x_{1i}\}}+\mathrm{o}_{p}(n^{-1/2}).
\end{align*}
Thus,
\begin{align*}
\mathrm{Rm}(\mathbf{w})&=\mathrm{Rm}_{2}(\mathbf{w})+\mathrm{o}_{p}(n^{-1/2})\\
 &= \binom n2 ^{-1}\sum_{i\neq j} 2^{-1}(\zeta_{ij}+\zeta_{ji})+\mathrm{o}_{p}(n^{-1/2}),
\end{align*}
where
\begin{align*}
\zeta_{ij}=\delta_{i}m(x_{i})\exp(\gamma y_{i})\mathrm{K}_{h}(x_{1i},x_{1j})\frac{1-\delta_{j}-\exp\{-g_{\gamma}(x_{1i})\}\delta_{j} \exp(\gamma y_{j})}{E\{\delta \exp(\gamma Y)\mid x_{1i}\}}.
\end{align*}
From the theory of U-statistics \citep{van}, we have 
\begin{align*}
\mathrm{Rm}(\mathbf{w})&=\frac{1}{n}\sum_{i=1}^{n}\left\{E(\zeta_{ij}|w_{i})+E(\zeta_{ji}|w_{i})\right\}+\mathrm{o}_{p}(n^{-1/2})\\    
&=\frac{1}{n}\sum_{i=1}^{n}E\{\delta_{i}m(x_{i})\exp(\gamma y_{i})\mid x_{1i}\}\frac{1-\delta_{i}-\exp\{-g_{\gamma}(x_{1i})\}\delta_{i} \exp(\gamma y_{i})}{E\{\delta \exp(\gamma Y)\mid x_{1i}\}}+\mathrm{o}_{p}(n^{-1/2})\\
&= \frac{1}{n}\sum_{i=1}^{n}\frac{E\{\delta m(X)\exp(\gamma Y)\mid x_{1i}\}}{E\{\delta \exp(\gamma Y)\mid x_{1i}\}}\left\{1-\frac{\delta_{i}}{\pi(x_{1i},y_{i};g_{\gamma},\gamma)}\right\}+\mathrm{o}_{p}(n^{-1/2}).
\end{align*}
From the first line to the second line, we use the fact $E(\zeta_{ij}|w_{i})=0$ and the order of the bias term of the nonparametric estimator is  $\mathrm{o}_{p}(n^{-1/2})$.  
By plugging the above $\mathrm{Rm}(\mathbf{w})$ in \eqref{eq:residual} , we have
\begin{align*}
    0 &=  \frac{1}{n}\sum_{i=1}^{n}\left\{\frac{\delta_{i}}{\pi_{i}(x_{1i},y_{i};\hat{g}_{\gamma},\gamma)}-1\right\}m(x_{i})\\
    &= \frac{1}{n}\sum_{i=1}^{n}\left\{\frac{\delta_{i}}{\pi_{i}(x_{1i},y_{i};g_{\gamma},\gamma)}-1\right\}\left [m(x_{i})-\frac{E\{\delta\exp(\gamma Y)m(X)\mid x_{1i}\}}{E\{\delta \exp(\gamma Y)\mid x_{1i}\}}\right]+\mathrm{o}_{p}(n^{-1/2}).
\end{align*}
Therefore, the asymptotic variance is calculated as $A^{-1}BA^{-1}$:
\begin{align*}
    A &= \nabla_{\gamma}E\left(\frac{1}{n}\sum_{i=1}^{n}\left\{\frac{\delta}{\pi(X_{1},Y;g_{\gamma},\gamma)}-1\right\}\left[m(X)-\frac{E\{\delta\exp(\gamma Y)m(X)\mid X_{1}\}}{E\{\delta \exp(\gamma Y)\mid X_{1}\}}\right]\right)\mid _{\gamma^{*}} \\
    &= E\left(\frac{1}{n}\sum_{i=1}^{n}\nabla_{\gamma}\left\{\frac{\delta}{\pi(X_{1},Y;g_{\gamma},\gamma)}\right\}\left [m(X)-\frac{E\{\delta\exp(\gamma Y)m(X)\mid X_{1}\}}{E\{\delta \exp(\gamma Y)\mid X_{1}\}}\right]\right)\mid _{\gamma^{*}} \\
    &= E\left(\frac{1-\pi}{\pi}\pi\left \{Y-E(Y  \mid X_{1},\delta=0)\right\}\left[m(X)-E\{m(X)\mid X_{1},\delta=0\}\right]\right)\mid _{g^{*},\gamma^{*}}, \\
    B &= E\left(\left\{\frac{\delta}{\pi(X_{1},Y;g_{\gamma},\gamma)}-1\right\}^{2}\left [m(X)-\frac{E\{\delta\exp(\gamma Y)m(X)\mid X_{1}\}}{E\{\delta \exp(\gamma Y)\mid X_{1}\}}\right]^{2}\right)\mid _{\gamma^{*}} \\
    &=  E\left(E\left[\left\{\frac{\delta}{\pi(X_{1},Y;g_{\gamma},\gamma)}-1\right\}^{2}|X\right]\left [m(X)-\frac{E\{\delta\exp(\gamma Y)m(X)\mid X_{1}\}}{E\{\delta \exp(\gamma Y)\mid X_{1}\}}\right]^{2}\right)\mid _{\gamma^{*}} \\
    &= E\left(\frac{1-\pi}{\pi}\left[m(X)-E\{m(X)\mid X_{1},\delta=0\}\right]^{2}\right)\mid _{g^{*},\gamma^{*}}.
\end{align*}
Note that from the first line to the second line, the calibration condition is used. From the second line to the third line, 
we used 
\begin{align*}
    &\nabla_{\gamma}\left\{\frac{1}{\pi(x_{1},y;g_{\gamma},\gamma)}\right\}|_{\gamma^{*}}\\
    &= \left
    [\exp(-g_{\gamma}+\gamma y)y-\exp(-g_{\gamma}+\gamma y)\frac{E\{\delta Y\exp(\gamma Y)\mid x_{1}\}}{ E\{\delta \exp(\gamma Y)\mid x_{1}\}}\right]|_{\gamma^{*}} \\
    &= \left[\frac{1-\pi}{\pi}\{y-E(Y  \mid x_{1},\delta=0)\}\right]|_{g^{*},\gamma^{*}}
\end{align*}

\subsection{Proof of Theorem 3}

We can replace the conditional expectation including the nonparametric component with the expectation as follows because of the calibration condition. This type of reasoning is already known in the literature. For example,  see Chapter 10.2 in \cite{TsiatisAnastasiosA.AnastasiosAthanasios2006Stam} and page $408$ in \cite{BolthausenErwin2002LoPT}. Thus, we obtain 
\begin{align}
&\frac{1}{n}\sum_{i=1}^{n}\left\{\frac{\delta_{i}}{\pi_{i}(x_{1i},y_{i};\hat{g}_{\gamma},\gamma)}-1\right\}\frac{\mathrm{\tilde{E}
    }\{\delta \exp(\gamma Y)\pi(X_{1},Y;\hat{g}_{\gamma},\gamma)Y |x_{i}\}}{\mathrm{\tilde{E}}\{\delta \exp(\gamma Y)\mid x_{i}\}} \nonumber \\
 =&\frac{1}{n}\sum_{i=1}^{n}\left\{\frac{\delta_{i}}{\pi_{i}(x_{1i},y_{i};\hat{g}_{\gamma},\gamma)}-1\right\}\frac{E\{\delta \exp(\gamma Y)\pi(X_{1},Y;\hat{g}_{\gamma},\gamma)Y  \mid x_{i}\}}{E\{\delta \exp(\gamma Y)\mid x_{i}\}} +\mathrm{o}_{p}(n^{-1/2}) \label{eq:p-ml-2}. 
\end{align} 
If we substitute $\hat{g}_{\gamma}(x_{1})$ in the conditional expectation with a linearlized estimator as in the proof of Theorem 2.1, the expression \eqref{eq:p-ml-2} is equal to 
\begin{align*}
&\frac{1}{n}\sum_{i=1}^{n}\left\{\frac{\delta_{i}}{\pi_{i}(x_{1i},y_{i};\hat{g}_{\gamma},\gamma)}-1\right\}\frac{E\{\delta \exp(\gamma Y)\pi(X_{1},Y;g_{\gamma},\gamma)Y |x_{i}\}}{E\{\delta \exp(\gamma Y)\mid x_{i}\}}  +\mathrm{o}_{p}(n^{-1/2}).
\end{align*}
Then, as in the proof of Theorem 1, the above term is equal to 
\begin{align*}
\frac{1}{n}\sum_{i=1}^{n}\left\{\frac{\delta_{i}}{\pi_{i}(x_{1i},y_{i};g(x_{1}),\gamma)}-1\right\}\left [q(x_{i})-\frac{E\{\delta\exp(\gamma Y)q(X)\mid x_{1i}\}}{E\{\delta \exp(\gamma Y)\mid x_{1i}\}}\right]+\mathrm{o}_{p}(n^{-1/2}),
\end{align*}
where 
\begin{align*}
    q(x)=\frac{E\{\delta \exp(\gamma Y)\pi(X_{1},Y;g_{\gamma},\gamma)Y |x\}}{E\{\delta \exp(\gamma Y)\mid x\}}.
\end{align*}
Then, the asymptotic variance is obtained as $A^{-1}BA^{-1}$:
\begin{align*}
    A &= \nabla_{\gamma}E\left(\frac{1}{n}\sum_{i=1}^{n}\left\{\frac{\delta_{i}}{\pi_{i}(X_{1i},y_{i};g_{\gamma},\gamma)}-1\right\}\left [q(X_{i})-\frac{E\{\delta\exp(\gamma y)q(x_{i})\mid x_{1i}\}}{E\{\delta \exp(\gamma y)\mid X_{1i}\}}\right]\right)\mid _{\gamma^{*}}\\
    &= E\left(\frac{1}{n}\sum_{i=1}^{n}\nabla_{\gamma}\left\{\frac{\delta_{i}}{\pi_{i}(X_{1i},Y_{i};g_{\gamma},\gamma)}\right\}\left \{q(X_{i})-\frac{E\{\delta\exp(\gamma y)q(X_{i})\mid X_{1i}\}}{E\{\delta \exp(\gamma Y)\mid X_{1i}\}}\right\}\right)\mid _{\gamma^{*}} \\
    &= E\left (O(X,Y) \pi \left\{Y-E_{0}\left (Y  \mid X\right)\right\}\left[E_{0}\{\pi Y  \mid X\}-E_{0}\{\pi(X,Y) Y  \mid X_{1}\}\right]\right)\mid _{g^{*},\gamma^{*}} ,\\
    B &= E\left (O(X,Y) \left[E_{0}\{\pi Y  \mid X\}-E_{0}\{\pi(X,Y;g,\gamma) Y  \mid X_{1}\}\right]^{2}\right)\mid _{g^{*},\gamma^{*}}.
\end{align*}
From the second line to the third line, we use 
\begin{align*}
    E\{E(\pi Y  \mid X,\delta=0)\mid X_{1},\delta=0\}= E(\pi Y  \mid X_{1},\delta=0).
\end{align*}

\subsection{Proof of Lemma 3.1 and Theorem 1}

We have 
\begin{align*}
0 &= \sum_{i=1}^{n}\delta_{i}\left\{1-\pi(x_{1i},y_{i};\hat{g}_{\gamma},\gamma)\right\}y_{i}+(1-\delta_{i})\frac{\tilde{E}\{-\delta\exp(\gamma Y)\pi(X_{1},Y;\hat{g}_{\gamma},\gamma))Y  \mid x_{i}\}}{\tilde{E}\{\delta\exp(\gamma Y)\mid x_{i}\}} \\
&=\frac{1}{n}\sum_{i=1}^{n}\left\{\frac{\delta_{i}}{\pi_{i}(x_{1i},y_{i};\hat{g}_{\gamma},\gamma)}-1\right\}\frac{E\{\delta \exp(\gamma Y)\pi(X_{1},Y;\hat{g}_{\gamma},\gamma)Y |x_{i}\}}{E\{\delta \exp(\gamma Y)\mid x_{i}\}}+\mathrm{o}_{p}(n^{-1/2}).
\end{align*}
From the first line to the second line, we use a result in the proof of \cite{morikawa17}.  This concludes the proof of Lemma 3.1. For the rest of the proof, we leave it to the proof of Theorem 3.

\subsection{Proof of Lemma 3.2}

When $g(x_{1})$ is known, the asymptotic variance is written as $n^{-1}A^{-1}BA^{-1}$:
\begin{align*}
    A &= E\left\{O(X,Y)\pi Y m(X)\right\}|_{\gamma^{*}}= E\left[E\left\{O(X,Y)\pi Y  \mid X \right\}m(X)\right]|_{\gamma^{*}},\\
    B &= E\left[E\left\{O(X,Y)\mid X\right\}m^{2}(X)\right]|_{\gamma^{*}}.
\end{align*}
We can use the Cauchy Schwartz inequality for the expectation:
\begin{align*}
    A^{-1}BA^{-1}\geq  E\left[E\left\{O(X,Y)\mid X\right\}^{-1}E\left\{O(X,Y)\pi Y  \mid X \right\}^{2}\right]|_{\gamma^{*}}
\end{align*}
The equality holds when $m(X;\gamma)\mid _{\gamma^{*}}$ is a proportional to 
\begin{align*}
    E\left\{O(X,Y)\mid X\right\}^{-1}E\left\{O(X,Y)\pi Y  \mid X \right\}|_{\gamma^{*}}. 
\end{align*}
This concludes the proof. 

\subsection{Proof of Lemma 3.3}

For the former statement, we can prove similarly as in Theorem 3. For the latter statement, it is proved because estimators are reduced to the form of $\hat{\gamma}_{p-gmm}$ when the outcome model is misspecified. 

\subsection{Proof of Theorem 4}

First, we show the asymptotic form of $\hat{\mu}_{ipw}$. Although a similar result is obtained in \cite{wang14}, our result is slightly different. The difference is that $E(y  \mid x_{1},\delta=0)$, which appear in the terms of $B_{1}$ and $B_{2}$, is $E(y  \mid x_{1})$ in their result. 

The estimator can be expanded as:
\begin{align*}
\hat{\mu}_{ipw}&=  \frac{1}{n}\sum_{i=1}^{n}\delta_{i}y_{i}\left[1+\exp\{-\hat{g}_{\hat{\gamma}}(x_{1i})\}+\hat{\gamma} y_{i}\right]\\
&= J_{1}+J_{2}+J_{3}+\mathrm{o}_{p}(n^{-1/2}),
\end{align*}
where
\begin{align*}
    J_{1} &= \frac{1}{n}\sum_{i=1}^{n}\frac{\delta_{i}y_{i}}{\pi(x_{i},y_{i};g^{*},\gamma^{*})}\\
   J_{2} &= \frac{1}{n}\sum_{i=1}^{n}\delta_{i}y_{i}\exp(\hat{\gamma} y_{i})\left[\exp\{-\hat{g}_{\gamma}(x_{1i})\}-\exp\{-g_{\hat{\gamma}}(x_{i})\}\right] \\
    J_{3} &= \frac{1}{n}\sum_{i=1}^{n}\delta_{i}y_{i}\left[\exp\{-g_{\hat{\gamma}}(x_{i})+\hat{\gamma}y_{i}\}-\exp\{-g_{\gamma^{*}}(x_{i})+\gamma^{*} y_{i}\}\right].
\end{align*}
As in the proof of Theorem 2.1, the term $J_{2}$ is 
\begin{align*}
    J_{2} &= \frac{1}{n}\sum_{i=1}^{n}\frac{E\{\delta \exp(\hat{\gamma} Y)Y \mid x_{1i}\}}{E\{\delta \exp(\hat{\gamma} 
    Y)\mid x_{1i}\}}\left\{1-\frac{\delta_{i}}{\pi(x_{1i},y_{i};g_{\hat{\gamma}}(x_{1}),\hat{\gamma})}\right\}+\mathrm{o}_{p}(n^{-1/2})\\
    &= \frac{1}{n}\sum_{i=1}^{n}\frac{E\{\delta \exp(\gamma Y)Y  \mid x_{1i}\}}{E\{\delta \exp(\gamma Y)\mid x_{1i}\}}|_{\gamma^{*}}\left\{1-\frac{\delta_{i}}{\pi(x_{1i},y_{i};g_{\hat{\gamma}}(x_{1}),\hat{\gamma})}\right\}+\mathrm{o}_{p}(n^{-1/2})\\
    &= \frac{1}{n}\sum_{i=1}^{n}E(Y  \mid x_{1i},\delta=0)\left\{1-\frac{\delta_{i}}{\pi(x_{1i},y_{i};g_{\hat{\gamma}}(x_{1}),\hat{\gamma})}\right\}+\mathrm{o}_{p}(n^{-1/2})\\
    &=\frac{1}{n}\sum_{i=1}^{n}E(Y  \mid x_{1i},\delta=0)\left\{1-\frac{\delta_{i}}{\pi(x_{1i},y_{i};g^{*}(x_{1}),\gamma^{*})}\right\}+\mathrm{o}_{p}(n^{-1/2}) \\
    &-\frac{1}{n}\sum_{i=1}^{n}\delta_{i}E(Y  \mid x_{1i},\delta=0)\exp\{-g_{\gamma^{*}}(x_{i})\}\exp(\gamma^{*} y_{i})\{y_{i}-E(Y  \mid x_{1i},\delta=0)\}(\hat{\gamma}-\gamma^{*})
\end{align*}
From the first line to the second line, we have used a calibration condition. From the third line to the fourth line, we have used a delta method. Similarly, from the delta method, the term $J_{3}$ is 
\begin{align*}
    J_{3} = \frac{1}{n}\sum_{i=1}^{n}\delta_{i}y_{i}\exp\{-g_{\gamma^{*}}(x_{i})\}\exp(\gamma^{*} y_{i})\{y_{i}-E(Y  \mid x_{1i},\delta=0)\}(\hat{\gamma}-\gamma^{*})+\mathrm{o}_{p}(n^{-1/2}). 
\end{align*}

Organizing the above result, we obtain
\begin{align*}
\hat{\mu}_{ipw}&=C_{1}+C_{2}+C_{3}+\mathrm{o}_{p}(n^{-1/2}), \\
    C_{1}&=\frac{1}{n}\sum_{i=1}^{n}E(Y  \mid x_{1i},\delta=0),\\
    C_{2}&=\frac{1}{n}\sum_{i=1}^{n}\frac{\delta_{i}}{\pi(x_{i},y_{i};g_{\gamma^{*}},\gamma^{*})}\{y_{i}-E(Y  \mid x_{1i},\delta=0)\}, \\
    C'_{3}&= H'_{1}(\hat{\gamma}-\gamma^{*}),
\end{align*}
where 
\begin{align*}
    H_{1}' &= \frac{1}{n}\sum_{i=1}^{n}\delta_{i}\exp\{-g^{*}(x_{i})+\gamma^{*}y_{i}\}\{y_{i}-E(y  \mid x_{1i},\delta=0)\}^{2}\\
    &= \frac{1}{n}\sum_{i=1}^{n}\delta_{i}\left(\frac{1}{\pi_{i}}-1\right)\mid _{g^{*},\gamma^{*}}\{y_{i}-E(y  \mid x_{1i},\delta=0)\}^{2}.
\end{align*}
From Slutsky's theorem, we have 
\begin{align*}
C'_{3}=E[(1-\pi)\{Y-E[Y \mid x_{1i},\delta=0\}^{2}]|_{g^{*},\gamma^{*}}(\hat{\gamma}-\gamma^{*})+\mathrm{o}_{p}(n^{-1/2}).
\end{align*}
This concludes the proof.

Next, we show the asymptotic form of $\hat{\mu}_{mp}$. From Theorem 1 in \cite{Kim11}, the following is yielded:

\begin{align}
\hat{\mu}_{mp} &= \frac{1}{n}\sum_{i=1}^{n}\frac{\delta_{i}y_{i}}{\pi(x_{i};g_{\hat{\gamma}},\hat{\gamma})}+\left\{1-\frac{\delta_{i}}{\pi(x_{i};g_{\hat{\gamma}},\hat{\gamma})}\right\} \frac{E(\delta \exp(\gamma^{*} ) Y  \mid x_{i})}{E(\delta \exp(\gamma^{*} Y)\mid x_{i})}+\mathrm{o}_{p}(n^{-1/2}) \nonumber \\
&= \frac{1}{n}\sum_{i=1}^{n}\frac{\delta_{i}}{\pi(x_{i};g_{\hat{\gamma}},\hat{\gamma})}\left\{y_{i}-E(Y  \mid x_{i},\delta=0)\right\}+E(Y  \mid x_{i},\delta=0)+\mathrm{o}_{p}(n^{-1/2})\label{eq:mu_mp} \\
&= \frac{1}{n}\sum_{i=1}^{n}\frac{\delta_{i}}{\pi(x_{i};g_{\gamma^{*}},\gamma^{*})}\left\{y_{i}-E[Y  \mid x_{i},\delta=0]\right\}+E(y|x_{i},\delta=0)+\mathrm{o}_{p}(n^{-1/2}) \nonumber \\
&+ H_{2}'(\hat{\gamma}-\gamma^{*}) \nonumber,
\end{align}
where 
\begin{align*}
H_{2}' = \frac{1}{n}\sum_{i=1}^{n}\frac{\delta_{i}\{1-\pi(x_{i})\}}{\pi(x_{i})}|_{g^{*},\gamma^{*}}\left\{y_{i}-E(Y \mid x_{i},\delta=0)\right\}\left\{y_{i}-E(Y \mid x_{1i},\delta=0)\right\}.
\end{align*}
From the second line to the third line, we use the delta method. In addition, we have 
{\small
\begin{align*}
    &E\left[\frac{\delta_{i}\{1-\pi(x_{i})\}}{\pi(x_{i})}|_{g^{*},\gamma^{*}}\left(y_{i}-E(Y \mid X_{i},\delta=0)\right)\left\{y_{i}-E(Y \mid X_{1i},\delta=0)\right\}\right]\\
    =& E\left[(1-\delta)\{Y-E(Y \mid X,\delta=0)\}\{Y-E(Y \mid X,\delta=0)+E(Y \mid X,\delta=0)-E(Y \mid X_{1},\delta=0)\}\right] \\
    =& E\left[\{Y-E(Y \mid X,\delta=0)\}\{Y-E(Y \mid X,\delta=0)+E(Y \mid X,\delta=0)-E(Y \mid X_{1},\delta=0)\}|\delta=0\right]P(\delta=0)\\
    =& E\left[\{Y-E(Y \mid X,\delta=0])^{2}\}|\delta=0\right]P(\delta=0) \\
    =& E\left[(1-\delta)\{Y-E(Y \mid X,\delta=0)\}^{2}\right]=H_{2}. 
\end{align*}
}
Slutsky's theorem concludes the proof.

Next, we show the result of $\hat{\mu}_{db}$. The estimator $\hat{\mu}_{db}$ is expanded as $\hat{\mu}_{mp}$: 
{\small
\begin{align*}
\hat{\mu}_{db}&=  \frac{1}{n}\sum_{i=1}^{n}\frac{\delta_{i}y_{i}}{\pi(x_{i};\hat{g}_{\hat{\gamma}},\hat{\gamma})}+\left\{1-\frac{\delta_{i}}{\pi(x_{i};\hat{g}_{\hat{\gamma}},\hat{\gamma})}\right \}\frac{\tilde{E}\{\delta \exp(\hat{\gamma} Y) Y \mid x_{i}\}}{\tilde{E}[\delta \exp(\hat{\gamma} Y)\mid x_{i}]}\\
&= \frac{1}{n}\sum_{i=1}^{n}\frac{\delta_{i}y_{i}}{\pi(x_{i};\hat{g}_{\hat{\gamma}},\hat{\gamma})}+\left\{1-\frac{\delta_{i}}{\pi(x_{i};\hat{g}_{\hat{\gamma}},\hat{\gamma})}\right\} \frac{E\{\delta \exp(\gamma^{*} Y) Y \mid x_{i}\}}{E\{\delta \exp(\gamma^{*} Y)\mid x_{i}\}}+\mathrm{o}_{p}(n^{-1/2})\\
&=\frac{1}{n}\sum_{i=1}^{n}\frac{\delta_{i}}{\pi(x_{i};\hat{g}_{\hat{\gamma}},\hat{\gamma})}\left\{y_{i}-E(Y \mid x_{i},\delta=0)\right\}+E(Y \mid x_{i},\delta=0)+\mathrm{o}_{p}(n^{-1/2}) \\
&= \frac{1}{n}\sum_{i=1}^{n}\frac{\delta_{i}}{\pi(x_{i};g_{\hat{\gamma}},\hat{\gamma})}\left\{y_{i}-E(Y \mid x_{i},\delta=0)-E(Y \mid x_{1i},\delta=0)+E(E(Y \mid x_{i},\delta=0)\mid x_{1i},\delta=0))\right \}
\\&+E(Y \mid x_{i},\delta=0)+\mathrm{o}_{p}(n^{-1/2}) \\
&=\frac{1}{n}\sum_{i=1}^{n}\frac{\delta_{i}}{\pi(x_{i};g_{\hat{\gamma}},\hat{\gamma})}\left\{y_{i}-E(Y \mid x_{i},\delta=0)\right\}+\mathrm{o}_{p}(n^{-1/2}).
\end{align*}
}
From the first line to the second line, the  calibration condition is used to replace the  nonparametirc estimator with a true distribution. From the third line to the fourth line, we use the same argument as in Theorem 1 to replace $\hat{g}_{\hat{\gamma}}$ with $g_{\hat{\gamma}}$. The final line is the same expression as \eqref{eq:mu_mp}. This concludes the proof. 

\subsection{Proof of Lemma 4.1}

For the former statement, we can use the same argument as in the proof of Theorem 4. For the latter statement, when the working model is mis-specified, there exists $s(x)$ satisfying
\begin{align*}
    \frac{\tilde{E}\{\delta \exp(\hat{\gamma} Y) Y \mid x;\hat{\beta}\}}{\tilde{E}\{\delta \exp(\hat{\gamma} Y)\mid x;\hat{\beta}\}}=
    s(x;\hat{\gamma})+\mathrm{o}_{p}(1). 
\end{align*}
Then, we have 
\begin{align*}
\hat{\mu}_{db}&=  \frac{1}{n}\sum_{i=1}^{n}\frac{\delta_{i}y_{i}}{\pi(x_{i};\hat{g}_{\gamma},\hat{\gamma})}+\left\{1-\frac{\delta_{i}}{\pi(x_{i};\hat{g}_{\gamma},\hat{\gamma})}\right\}\frac{\tilde{E}\{\delta \exp(\hat{\gamma} Y) Y \mid x_{i}\}}{\tilde{E}\{\delta \exp(\hat{\gamma} Y)\mid x_{i}\}}\\
&= \frac{1}{n}\sum_{i=1}^{n}\frac{\delta_{i}y_{i}}{\pi(x_{i};\hat{g}_{\gamma},\hat{\gamma})}+\left\{1-\frac{\delta_{i}}{\pi(x_{i};\hat{g}_{\gamma},\hat{\gamma})}\right\}s(x_{i};\hat{\gamma})+\mathrm{o}_{p}(1) \\
&= \frac{1}{n}\sum_{i=1}^{n}\frac{\delta_{i}y_{i}}{\pi(x_{i};\hat{g}_{\gamma},\hat{\gamma})}+\mathrm{o}_{p}(1)=\hat{\mu}_{ipw}+\mathrm{o}_{p}(1).
\end{align*}
This concludes the proof. 

\section{Interpretation of \texorpdfstring{$\hat{\gamma}_{p-score}$}{a} from the perspective of maximizing observed likelihood}

By maximizing $E\{l(W;\hat{g}_{\gamma},\gamma)\mid W_{obs}\}$, where $l(W;\hat{g}_{\gamma},\gamma)$ denotes a full-likelihood, we can derive the following EM-type iterative estimator. 
\begin{itemize}
    \item E-step: calculate an approximation of an expected log-likelihood $E\{\log l(W)\mid W_{obs}\}$ as 
        \begin{align*}
         L(\gamma) = \sum_{i=1}^{n} \delta_{i}\log \pi(x_{1i},y_{i};\hat{g}_{\gamma},\gamma)+(1-\delta_{i})h_{\hat{\gamma}_{t}}(x_{i};\gamma),
        \end{align*} 
        where
        \begin{align*}
            h_{\hat{\gamma}_{t}}(x;\gamma)=\frac{\tilde{E}\{\delta \exp(\hat{\gamma}_{t}Y)\log(1-\pi(X,Y;\hat{g}_{\gamma},\gamma))\mid x\} }{\tilde{E}\{\delta\exp(\hat{\gamma}_{t}Y)\mid x\}}.
        \end{align*}.
\item M-step: maximize the above function with respect to $\gamma$. Update the value as $\hat{\gamma}_{t+1}$.
\item By iterating E-step and M-step, we define the converge point as an estimator $\hat{\gamma}_{p-mle}$.
\end{itemize}

The estimator $\hat{\gamma}_{p-mle}$ is essentially equal to $\hat{\gamma}_{p-score}$ because of the following reason. The estimator $\hat{\gamma}_{p-score}$ solves the equation $E\{\nabla_{\gamma}\log l(W;\hat{g}_{\gamma},\gamma)\mid W_{obs}\}=0$. It can be converted into an EM-style method. From this viewpoint, when $Y_{mis}$ is imputed by $W|W_{obs}$ in the E-step, solving the score equation $\nabla_{\gamma}\log l(W;\hat{g}_{\gamma},\gamma)=0$ in $\hat{\gamma}_{p-score}$ is equivalent to maximizing the c
omplete-data profile log-likelihood $\log l(W;\hat{g}_{\gamma},\gamma)$ with respect to $\gamma$ in $\hat{\gamma}_{p-mle}$.

\section{Consistency of estimators}

For simplicity, we discuss the consistency of the estimator $\hat{\gamma}_{p-gmm}$ based on the expression (7). For the consistency of other estimators, we can prove similarly. 
We define 
\begin{align*}
\hat{m}(w;\gamma)=\left \{\frac{\delta}{\pi_{p}(x,y;\gamma)}-1\right \}h(x;\gamma), \,m(w;\gamma)=\left\{\frac{\delta}{\pi(x,y;g_{\gamma},\gamma)}-1\right \}h(x;\gamma).
\end{align*}

For consistency of estimators based on profile semparametric estimators, three conditions are needed  \citep{BolthausenErwin2002LoPT}; (1) there exists a Glivenko-Cantelli class $F$ of functions with integrable envelope such that $\mathrm{P}(\{\hat{m}(w;\gamma)\}\in F)\to 1$, (2) $\sup_{\gamma\in \Gamma}|\hat{m}(w;\gamma)-m(w;\gamma)\mid \stackrel{p}{\rightarrow}0$ for all $x$, and (3) there exists a point $\gamma^{*} \in \Gamma$ such that $\inf_{|\gamma-\gamma^{*}|>\delta}E\{m(w;\gamma)\}>E\{m(w;\gamma)\}|_{\gamma^{*}}=0$ for every $\delta>0$. 

The first and second condition are technical conditions. These condition are reduced to more primitive conditions \citep{newey94}, which are immaterial for our work. The third condition is a common condition for the consistency of M-estimators without plug-in nonparametric estimators \citep{van}.

\centerline{ \textcolor{blue}{\bf REFERENCES} }
\bibliographystyle{chicago}
\bibliography{pfi}
\end{document}